\documentclass{amsart}
\usepackage{amsmath,amssymb,amsthm,amsfonts,amscd,mathrsfs}
\usepackage[all]{xy}

\usepackage{tikz-cd}
 \usepackage{animate}
 \usepackage{graphicx} 
 \usepackage{pb-diagram}
 \usepackage[]{todonotes}

\usepackage{pgf,tikz}
\usepackage{mathrsfs}
\usetikzlibrary{arrows}
\usepackage{changepage}
\usepackage{multirow}

\begin{document}
\vspace*{0.2in}

\begin{flushleft}
{\Large
\textbf\newline{Malaria intensity in Colombia by regions and populations} 
}
\newline
\\
Alejandro Feged-Rivadeneira\textsuperscript{1,2},
Andrés Ángel\textsuperscript{3},
Felipe González-Casabianca\textsuperscript{3},
Camilo Rivera\textsuperscript{4}
\\
\bigskip
\textbf{1} Department of Anthropology, Stanford University, Stanford, CA, USA
\\
\textbf{2} Department of Urban Management and Design, Universidad del Rosario, Bogotá, Colombia
\\
\textbf{3} Department of Mathematics, Universidad de los Andes, Bogotá, Colombia
\\
\textbf{4} Walmartlabs, Sunnyvale, CA, USA

\bigskip

%
%






\end{flushleft}
\section*{Abstract}
Determining the distribution of  disease prevalence among heterogeneous populations at the national scale is fundamental for  epidemiology and public health. Here, we use a combination of methods (spatial scan statistic, topological data analysis, epidemic profile) to study measurable differences in malaria intensity by regions and populations of Colombia. This study explores three main questions: What are the regions of Colombia where malaria is epidemic? What are the regions and populations in Colombia where malaria is endemic? What associations exist between epidemic outbreaks between regions in Colombia?  
\textit{Plasmodium falciparum} is most prevalent in the Pacific Coast, some regions of the Amazon Basin, and some regions of the Magdalena Basin. \textit{Plasmodium vivax}  is the most prevalent parasite in Colombia, particularly in the Northern Amazon Basin, the Caribbean, and municipalities of Sucre, Antioquia and Cordoba. Malaria has been reported to be most common among 15-45 year old men. We find that the age-class suffering high risk of malaria infection ranges from 20 to 30 with an acute peak at 25 years of age. Second, this pattern was not found to be generalizable across Colombian  populations, Indigenous and Afrocolombian populations experience endemic malaria (with household transmission). Third, clusters of epidemic malaria for \textit{Plasmodium vivax}  were detected across Southern Colombia  including the Amazon Basin and the Southern Pacific region.  \textit{Plasmodium falciparum}, was  is epidemic in 13 of the 1,123 municipalities  (1.2\%).  Some key locations act as bridges between epidemic and endemic regions. Finally, we generate a regional classification based on intensity and synchrony, dividing the country into epidemic areas and bridge areas.

\section{Introduction}

Malaria in Colombia has been studied from a variety of disciplines that describe disease patterns with  dimensions such as the diversity of the vector  \cite{rubio1997ecoregional,saenz2001mapas}, characteristics of the parasite \cite{carmona2004malaria}, social phenomena affecting disease transmission \cite{alexander2005case, arevalo2015clinical}, and geological phenomena  \cite{poveda2001coupling, gagnon2002nino}.  Mainly, national and local contexts are well understood for a country that presents unusual diversity of environments and social backgrounds (including vast cultural diversity), which, in turn, represents different characteristics of malaria transmission. In contrast with Sub-Saharan Africa, where malaria is commonly a deadly disease affecting  primarily  children, Colombia is not considered particularly relevant in malarial disease studies given the relatively low mortality when compare with Sub Saharan Africa. However, malaria in Colombia presents certain characteristics that resemble those observed in Southeast Asia. Colombia was one of the first countries where resistance to chloroquine-based treatment was reported. Varied malaria intensity among segregated and diverse populations inhabiting different and unique environments make Colombia one of the few cases where malaria is endemic and where disease patterns are inconsistent from regionally, in contrast to several countries that follow a consistent  pattern of infection, or whose segregated vulnerable populations do not differ in their epidemic patterns \cite{valero2006malaria,world2013malaria}. This does not mean that other countries  have a homogeneous experience of malaria intensity across subpopulations or regions. However, disease distribution among Colombian populations has caused the parasite to generate resistance to treatment, unlike several other countries in the world except for South East Asia.

Malaria is a complex disease, and factors associated to disease severity and resistance have been reported, yet genetic resistance to malaria is more understood than to any other human disease \cite{hill1992malaria}.  However, the strong geographical association between resistance to the pathogen and disease severity remains a major challenge to assess the causality of human genetic resistance \cite{hill1992malaria}. We know from evolutionary theory that  two critical factors for selection must occur: 1) a population with genetic diversity has to exist for selection to be able to operate; 2)   a differential in reproductive value of the trait in question for adaptation to evolve. Because African populations exhibit both genetic diversity and experience severe malaria, genetic resistance to the pathogen appears to have emerged independently in different foci \cite{greenwood1991some}. However, unlike Africa, Colombia has no record of human genetic resistance to malaria. On the contrary, the parasite appears to have developed resistance to treatment. Until recently, it was unknown whether pathogen resistance was the result of selection of mutant strains under drug pressure, the spread of resistant strains, or adaptation of previously sensitive parasites \cite{rosario1976genetics}. Genetic evidence suggests that resistant malaria emerged in at least 4 different geographical foci, consistent with the history of reports of resistant pathogens for \textit{Plasmodium falciparum} in the Thailand-Cambodia  border and Colombia in the 1950s, then spreading for two decades to South America, Asia and India, and then to Africa in Kenya and Tanzania in the late 1970s \cite{wellems2001chloroquine}. Resistant \textit{Plasmodium vivax}  was first reported in Papua-New Guinea in 1989, it is currently present in South East Asia, and suspected to occur in South America \cite{wellems2001chloroquine}. Studies have found resistant \textit{Plasmodium vivax} at a rate of 11\% in representative samples of all blood smears collected in two endemic geographical regions in Colombia: \textit{Llanos orientales} (Eastern Plains) and Urabá \cite{soto2001plasmodium}, while others have found no evidence of resistant \textit{Plasmodium vivax}  forms in the Pacific Coast and the Amazon Basin \cite{castillo2002assessment}. However, therapeutic failure rates of \textit{Plasmodium falciparum} (for representative samples of all blood smears collected) have been reported as high as 78\% for these same regions \cite{castillo2002assessment}, and 67\% in Antioquia \cite{blair2002resistance}. More resent assessments of malaria prevalence in endemic areas also suggest that uncomplicated malaria by low parasitemia is one of the biggest challenges for malaria control strategies \cite{arevalo2015clinical, vallejo2015high}, and studies indicate that the observed differences are not attributable to human genetic traits that confer resistance \cite{ortega2015evidence}.

Few studies have addressed the malarial epidemiology by regions and populations to explore the role of malaria intensity in the emergence of resistant forms of the parasite. However, the role of Colombia in the global epidemiological context of malaria indicates  that the country may present unique characteristics for disease transmission. Mainly, the presence, absence, and most importantly, emergence of resistant forms of the parasite in different regions suggests that isolated and distinct epidemic regions exist within the national boundaries, and such characteristics may play a distinctive role in the evolution of the parasite. Here we address  the intensity of malaria by regions and human populations in Colombia, and the degree that the epidemic characteristics between regions affect each other.

One key aspect remains poorly understood about malaria dynamics in Colombia: how many different epidemic regions exist, and how do subpopulations in these regions experience malaria.  During fieldwork, we interacted with local health officials who conducted  malaria prevention programs at both local and national levels. Each public health official had knowledge and expertise about epidemic dynamics in their specific territorial assignment. However, a lack of systematic approaches hamper the ability to formalize such knowledge.  Malaria intensity and the social aspects that condition the transmission of the parasite drive public health interventions.  However, the regional designations are  yet to be formalized based on analysis of malaria dynamics. Decisions about prevention strategies, and how to target the most vulnerable populations are made based primarily on the expertise of local health officials, as indicated by the classification by \cite{ruiz2006modelling, bouma1997predicting,poveda2001coupling}

Here we generate a systematic classification of the malarious regions and subpopulations of Colombia, to characterize locations and subpopulations with epidemiological aspects of the parasite. Here we address  three basic questions concerning/surrounding the intensity of malarial infection by ethnicity an region:

\begin{enumerate}
  \item Is this population experiencing higher malaria intensity than other regions of the  country?
  \item Is the parasite persisting endemically within this population?
  \item Are the epidemic characteristics of this subpopulation affecting other subpopulations?
\end{enumerate}

Specifically, we examine eight years of malarial case reports, they are examined for both  malaria intensity, synchrony and segregation by ethnicity. First, we employ  an outbreak detection algorithm \cite{kulldorff1997} widely used \cite{kulldorff1998evaluating,hjalmars1996childhood,burkom2003biosurveillance} to identify clusters in space with outbreaks of malaria. Second, we apply methods from Topological Data Analysis (TDA) \cite{mapperPBG,shapeofdata} to discover synchronous outbreaks: areas that present similar time patterns of malarial epidemic.  Finally, regional case reports are explored with descriptive statistics to analyze the intensity of malaria exposure by ethnicity.

\section{Background}

From  John Snow's seminal study of cholera in London, epidemiology has been a spatial discipline \cite{cameron1983john}.  Geographical disease patterns have been widely described for numerous pathogens and regions. We use three  methods to analyze malaria in Colombia: disease clustering, disease visualization, and ecological analysis.

The production  of good quality maps to understand and visualize risk of disease transmission is recognized as one of the fundamental tools for malaria control strategies, specifically, understanding the relationship between malaria endemicity and the health impact of malaria \cite{snow1996need}. Studies suggest that annual entomological inoculation rates (commonly computed as the product of the daily human biting rate, the sporozoite rates from the caught mosquitoes, and the days per year, 365 \cite{kilama2014estimating}) in Ghana (100-1000), Kenya (10-60) and Gambia (less than 10) are associated to prevention of all cause childhood mortality rates by insecticide treated bed nets, with efficacy of 17\%, 33\%, and 63\%, respectively \cite{snow1996need}. These results suggest that public health policies should vary according to malaria endemicity, since bed nets have been the linchpin of malaria prevention strategies since DDT was discontinued as a viable alternative. Yet, evidence suggests that there are several contexts in which bed nets are not efficient \cite{snow1996need}. In locations where malaria is intense, the use of bed nets is less efficient to prevent the burden of the disease.

Due to the scarcity of multi-sited studies across different countries, variation of the relationship between endemicity and overall health remains unknown \cite{snow1996need}. However, within country variation of malaria has been subject of numerous studies. One study that addresses such relationship is produced by \cite{omumbo1998mapping}, using GIS and malaria case reports to map malaria intensity in Kenya. Their results also question the use of treated bed nets in regions where malaria is intense, because in these communities, bednets are the most inefficient \cite{omumbo1998mapping}.

Spatial descriptions of variation in malarial infection within countries has been addressed using maps of risk of contracting the disease. For example, \cite{kleinschmidt2000} produced a more accurate visualization of risk of contracting malaria in Mali, by combining regression analysis with \textit{``krigging"} (i.e., an interpolation method similar to smoothing fitted values) to account for local responses to environmental conditions such as weather, population and other topographic and sociological features. Using those methods, they are able to identify regions where the risk is higher than represented in traditional maps \cite{kleinschmidt2000}. A similar approach, but based on entomological and demographic  geo-coded records, is implemented with a GIS analysis to describe local risk of infection based upon proximity to breeding sites and human populations \cite{kitron1994geographic}. \cite{beck1994remote} have  implemented a variation of these risk maps by integrating remote sensing data to identify locations of high transmission based on human-vector interaction for a region in Mexico, including variation by season. 

The applicability of mobile phone data to map human mobility with disease dynamics does pose some interesting caveats. First, the fraction of the population with high degree of mobility remains constant in some studies, but this does not necessarily mean that it is precisely that fraction of the population who is moving pathogens from one place to another \cite{candia2008uncovering}. The N\"ukak represent one of the most endemic and vulnerable populations in terms of malaria persistence, and are highly mobile. Furthermore, multi-scale network models of human mobility suggest that local migration plays an important role in the synchronization of epidemics among subpopulations \cite{balcana2009multiscale}, and suggests that small populations who are highly mobile play a fundamental role in the dispersal of epidemics.

Estimating the effect of migration on pathogen loads has been a growing interest of epidemiologists in the past years, and multiple methods have been implemented to address such interaction.  A different data-driven approach to examine the effect of human mobility on epidemics has been the gravity model, used to evaluate measles outbreaks, both by age-classes and by urban and rural settings \cite{bharti2008measles,ferrari2010episodic,ferrari2010rural}. The main finding of this approach was that population densities were the main driver of outbreak seasonality across different environments \cite{bharti2008measles,ferrari2010episodic,ferrari2010rural}. Furthermore, the same group has used nighttime light imagery to estimate the effect of changing patterns of population densities on disease outbreaks \cite{bharti2011explaining}. Unfortunately, few comparative studies exist to determine which method is more effective under which conditions and for which diseases. However, the method of nighttime light imagery provides good estimates of mobility of populations without access to phone services, often the most vulnerable populations in terms of disease prevalence.

Although the gravity model has been mostly used for directly transmitted diseases, understanding the effect of human mobility on disease epidemics, and more generally how disease disperses over space and time, has been one of the fundamental questions in contemporary spatial epidemiology. 

Two approaches have been used to analyze synchrony of disease outbreaks over space and time, controlling for seasonal and environmental variables. Spectral analysis has been used to describe the association between aggravation of asthma symptoms and temperature or atmospheric contamination levels  \cite{bishop1977statistical,cazelles2007time}, and the association between air pollution and mortality \cite{cazelles2007time}. This technique has been used to study the effect of climatic variation on cholera \cite{pascual2000cholera1,pascual2002cholera}, malarial epidemics \cite{pascual2006malaria}, and the seasonality of sexually transmitted diseases (STDs) \cite{grassly2005host}. 

Some authors suggest that these methods have limitations, because they can only be used for time-series data in which statistical proprieties do not change over time, yet, epidemic data are inherently complex and non-stationary \cite{cazelles2007time}. Furthermore, evidence suggests that epidemic data characteristics do change over time \cite{duncan1996whooping,rohani2003ecological, cazelles2007time}. 

The limitations of the spectral decomposition methods led to the implementation of the second technique that is most widespread in understanding disease dynamics over space and time, coupled with climatic and environmental conditions from a non-stationary perspective: wavelets, a method used to show how time-series vary as a function of time and space \cite{cazelles2007time, cazelles2014wavelet}.

Wavelet analysis has been used to study geographical hierarchies of measles epidemics, and the observed effect of vaccination policies over time \cite{grenfell2001travelling}. Associations between dengue epidemics and El Ni\~no Southern Oscillation (ENSO) have also been documented using this method \cite{cazelles2005nonstationary}. Kreppel\cite{kreppel2014non} found an association between ENSO, Indian Ocean Dipole (IOD) and plague dynamics in Madagascar, and \cite{onozuka2014effect} found similar effect of those two climatic phenomena on infectious gastroenteritis in Japan. \cite{morris2014complex} documented that Buruli ulcer is affected by short and long rainfall patterns in French Guiana, as well as stochastic events such as ENSO. The relationship between ENSO and cutaneous leishmaniasis has also been documented for Costa Rica \cite{chaves2006climate}. Jose\cite{jose2003scaling} have studied the changing patterns and seasonality of Australian rotavirus epidemics comparing a multiplicity of methods including wavelet analysis, and detected seasonal biannual and quinquennial periods, yet, a three year epidemic period was also found to be dominant. Spectral analysis confirms that serotype harmonics interact in a complex, non-linear fashion, yielding an observable overall pattern beyond the isolated dynamics of each separate serotype, that is more than the sum of the parts, and inherent dynamics remain unchanged but the amplitude of disease infection is modified \cite{jose2003scaling}.

Spatial analysis methods have been applied in disease cluster identification. The main approaches used are: K-cluster analysis,  detects global clusters based on each case point \cite{cuzick1990spatial}; the geographical machine \cite{openshaw1987mark} and the scan statistic \cite{kulldorff1997}, work by aggregating cases in different areas and performing a hypothesis test based on a Bernoulli null model, with the advantageous difference for the scan statistic that it can perform multiple tests simultaneously. We present an implementation of the scan statistic in this case study. Small, isolated outbreaks of malaria in specific communities have been identified as ``discrete mini epidemics", which represent  disease severity by using space-time cluster identification \cite{snow1993periodicity}. Disease risk by geographical location has also been implemented as simple logistic regressions that include altitude, and physical coordinates of each individual within a case-control study \cite{brooker2004spatial}. The scan statistic method has been used by \cite{coleman2009using} to identify disease clusters over space and time in a South African region. \cite{zhang2008spatial} have also implemented the scan statistic method in China to identify clusters and suggest public health resource optimization. Faires\cite{faires2014detection} used this method to identify clusters of \textit{Clostridium difficile} over time in Ontario, Canada. Duczma\cite{duczmal2015dry}  implemented the scan statistic to study Chagas' disease in Brazil, while \cite{occelli2014mapping} do the same for end-stage renal disease (ESRD) in northen France. In Virginia, the increasing burden of Lyme disease was documented using spatiotemporal scan statistics \cite{li2014spatial}. In all cases, studies were able to identify areas with more cases than expected, highlighting in many cases the relevance of regions that did not present a comparatively higher incidence. 

Globally, \cite{rogers2000global} have used maximum likelihood methods (i.e. a similar approach to the scan statistic) to predict areas where malaria is likely to expand as a result of climate change.

Topological Data Analysis (TDA) applies ideas from the mathematics area of topology to study high dimensional data by obtaining invariants and useful representations of the shape of the data. 

In recent years, TDA has been used to find subgroups of individuals with unique genetic and prognostic profiles of breast cancer \cite{shapeofdata,cancertda}, political alliances in the congress \cite{shapeofdata}, profiles of basketball players  \cite{shapeofdata}, pathogen persistence in soil \cite{soiltda}, novel patterns in spinal and brain injury \cite{spinaltda}, subgroups of individuals with different complications from type 2 diabetes \cite{diabetestda}, among others.

\section{Methods}

The analysis for this study was generated from case reports based on active and passive detection methods compiled by the Colombian government. Reports of malaria cases are mandatory. Treatment is provided for free to each case, and a positive test is required to disburse the medication. Data were accessed by requesting a user to query the Sispro database of reported cases by the Ministry of Health. 

\begin{figure}
\hspace*{-0.5in}
\centering
  \includegraphics{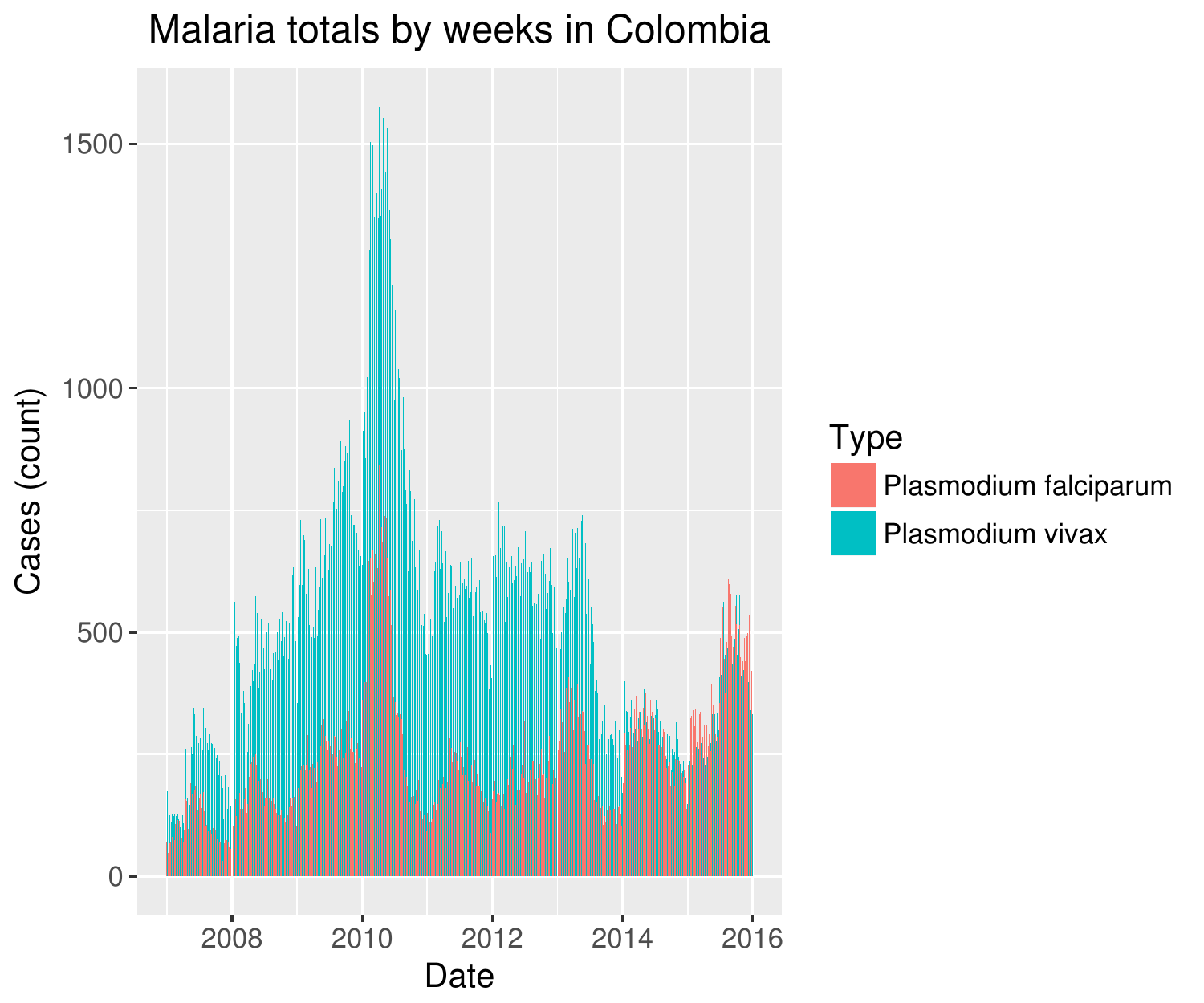}
    \caption{Distribution of registered malaria cases in Colombia, between the years 2007 and 2015 }
  \label{fig:cases_week}
\end{figure}

All cases are laboratory confirmed and geocoded to the municipality level. We included data for 1,156 municipalities, that range  in area from 15 to 65,674 km\textsuperscript{2}; total area sampled was 1,142 million km\textsuperscript{2}. For each municipality, we also analyzed ethnic membership, comprising 3,369 different populations. The distribution of malaria cases and segmentation of the data are displayed in Fig \ref{fig:cases_week} and Table \ref{table:summary}

\begin{table}
\centering
\caption{Summary and segmentation of registered malaria cases in Colombia between the years 2007 and 2015 }
\label{table:summary}
\begin{tabular}{|c|c|c|c|}
\hline
\multicolumn{2}{|c|}{Attribute}                          & Total Cases   & Percentage (\%)  \\ \hline
\multirow{2}{*}{Sex}             & M                     & 228075        & 63.32       \\ \cline{2-4} 
                                 & F                     & 132075        & 36.67       \\ \hline
\multirow{3}{*}{Ethnicity}       & Afro                  & 110333        & 30.63       \\ \cline{2-4} 
                                 & Indigenous            & 38199         & 10.60       \\ \cline{2-4} 
                                 & Other                 & 211618        & 58.75       \\ \hline
\multirow{2}{*}{Type of Malaria} & \textit{Plasmodium falciparum} & 115260        & 32.00       \\ \cline{2-4} 
                                 & \textit{Plasmodium vivax}      & 244890        & 67.99       \\ \hline
\multicolumn{2}{|c|}{Total}                              & \multicolumn{2}{c|}{360150} \\ \hline
\end{tabular}
\end{table}

\subsection{Clustering}

The two main objectives are to determine if malaria outbreaks exist in Colombia, and, if so, to determine their location. To address these objectives,  we apply scan statistics to perform a hypothesis test in each municipality, examining whether it presents an outbreak. These approaches have been widely used in epidemiological studies \cite{kulldorff1997}, \cite{neill2009}, and \cite{neill2009b}.

The model to test hypothesis is mainly based on the Bernoulli model of \cite{kulldorff1997} using the r-package Spatial-Epi\cite{kulldorf_package}. 

Given the data aggregated by municipality for 2007-2015, each record is assigned to the centroid of the municipality.  Because the set of possible outbreaks (all possible aggregations of neighboring municipalities) is almost unlimited in terms of shape and size, so the step is to approximate this set. In this case, a grid \textit{G} of \textit{N} by \textit{N} was overlaid onto Colombia's jurisdictional boundaries and then the set of possible outbreaks is limited to all the possible sub rectangles within the grid.

Now, under the Bernoulli model we consider a measurement \textit{m} for each rectangle \textit{R} contained in \textit{G}, where \textit{m(R)} corresponds to an integer and in our specific case, the number of individuals inside the given rectangle. This leads us to assume that there is a rectangle \textit{Z} contained in  \textit{G} such that each individual inside \textit{Z} has a probability \textit{p} of being infected, while the individuals outside \textit{Z} have a  probability \textit{q}. Let \textit{n\textsubscript{R}} be the number of observed malaria cases inside \textit{R}, so by assuming a Bernoulli and the following hypothesis for our unknown variables \textit{p} and \textit{q}:

\begin{eqnarray}
\label{eq:hipotesis}
H_0:& \; p = q \\
H_1:& \; p > q \,
\end{eqnarray}

we have these possible distributions:

\begin{itemize}
    \item Assuming \textit{H\textsubscript{0}}:
        \begin{eqnarray}
             n_R \sim Bin(m(R),p) \; \forall R \subseteq G
        \end{eqnarray}
    \item Assuming \textit{H\textsubscript{1}}:
        \begin{eqnarray}
             n_R \sim Bin(m(R),p) \; \forall R \subseteq Z \;\; \mbox{ and } \;\; n_R \sim Bin(m(R),q) \; \forall R \subseteq Z^C  
        \end{eqnarray}
\end{itemize}

And hence, under \textit{H\textsubscript{1}}, we have that \textit{Z} is a region with potential malaria outbreak.

Lastly, the third and final step is to establish a measure of density for each subrectangle, in this case the likelihood ratio. This measurement of density has desirable properties to compare different sized rectangles \cite{neill2009b}. \cite {kulldorff1997} derives the formula for likelihood ratio of a generic region. The scan statistic (\textit{ss}) is defined as the highest density measurements for all subrectangles:

\begin{eqnarray}
		ss^*  = \max_{R} \; ss(R)\\
\end{eqnarray}  
\begin{eqnarray}
		ss(R) = p^{n_R}(1-p)^{m(R)-n_R}q^{n_G - n_R}(1-q)^{(m(G) - m(R)) - (n_G - n_R)}
\end{eqnarray}

The local measurement \textit{ss(R)} can be interpreted as the likelihood that subrectangle \textit{R} is an outbreak.\\

To test the hypothesis represented in Eq~\ref{eq:hipotesis} a Monte Carlo simulation was used to obtain the histogram of the statistic  \textit{ss\textsuperscript{*}} under the null hypothesis. Finally, it assesses the value of  \textit{ss\textsuperscript{*}} with the observed data. If \textit{p} is greater than \textit{0.05} under the null model,  \textit{H\textsubscript{0}}  is rejected and we assume an outbreak.

\subsection{Synchronous epidemic visualization}

The main objectives are to determine whether abnormal behaviors are related across municipalities and finding, if a relationship exists, groups of them that have a similar temporal patterns, independent of their geographical layout. To address these matters we turn to Topological Data Analysis. 

We apply the method \textit{Mapper}, introduced in \cite{mapperPBG}, that constructs a representation of the data in the form a graph. This graph allows the analysis and visualization of the data. The vertices of the graph correspond to local clusters and the interactions between these clusters is enconded on the edges of the graph. 

\textit{Mapper} detects phenomena that appear both at large and small scale better than other  methods, such as principal component analysis (PCA) and cluster algorithms. \textit{Mapper} can be considered an hybrid method that is doing partial clustering, the regions where the clustering is done is guided by the filter. It is a refinement of clustering and scatterplot methods like PCA.

\begin{figure}
\hspace*{-2.2in}
\centering
  \includegraphics{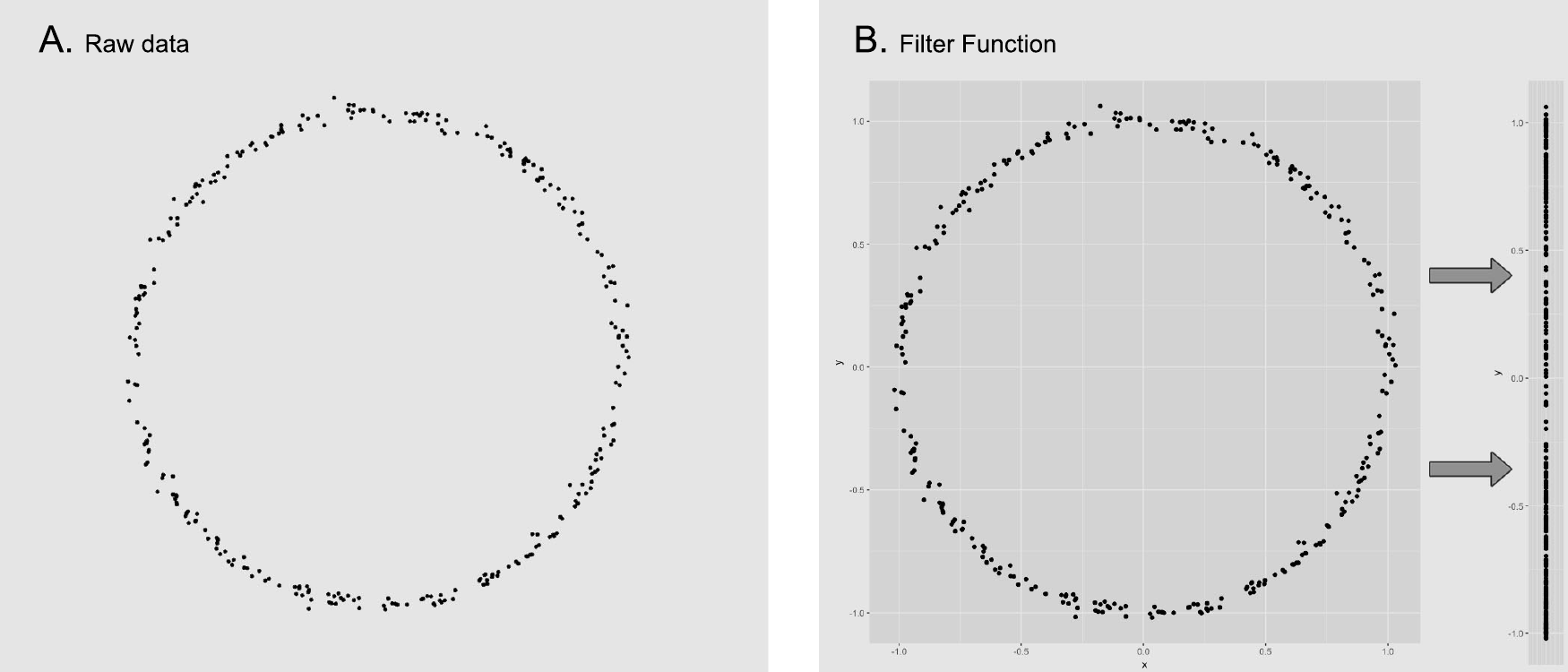}
    \caption{We start with a given data set (image A), for this example the points correspond to a sample of the unitary circle with a small amount of noise. For convenience we will use the euclidean distance to calculate the distance between each pair of points. In the next step, we  select the projection onto the \textit{Y} coordinate as our filter function and apply it to the data set (image B).   }
  \label{fig:mapper1}
\end{figure}

\begin{figure}
\hspace*{-2.2in}
\centering
  \includegraphics{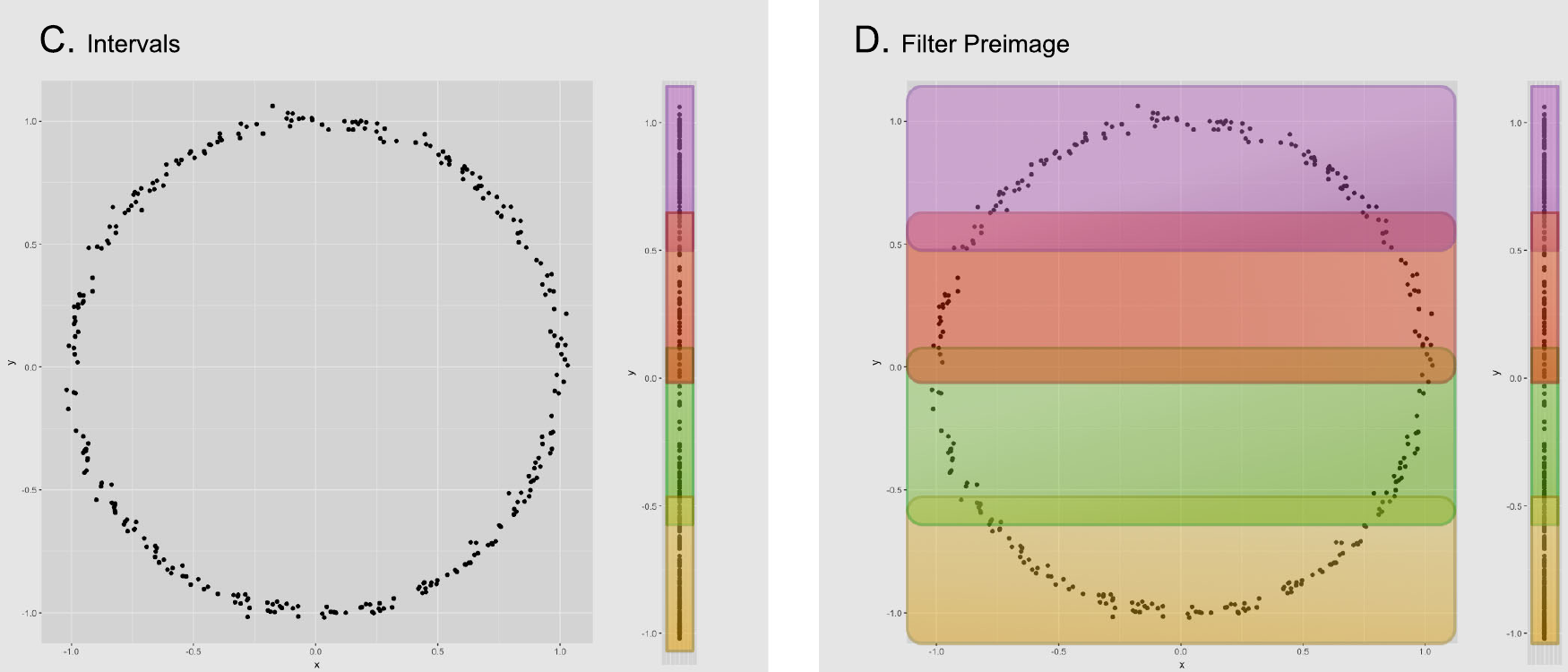}
    \caption{We now divide the image of the data set (under the filter function) into evenly distributed overlapping intervals (image C) and compute the corresponding points in their pre-image (image D). Notice how each pair of overlapping intervals, define two different subsets of data that can have elements in common.}
  \label{fig:mapper2}
\end{figure}

\begin{figure}
\hspace*{-2.2in}
\centering
  \includegraphics{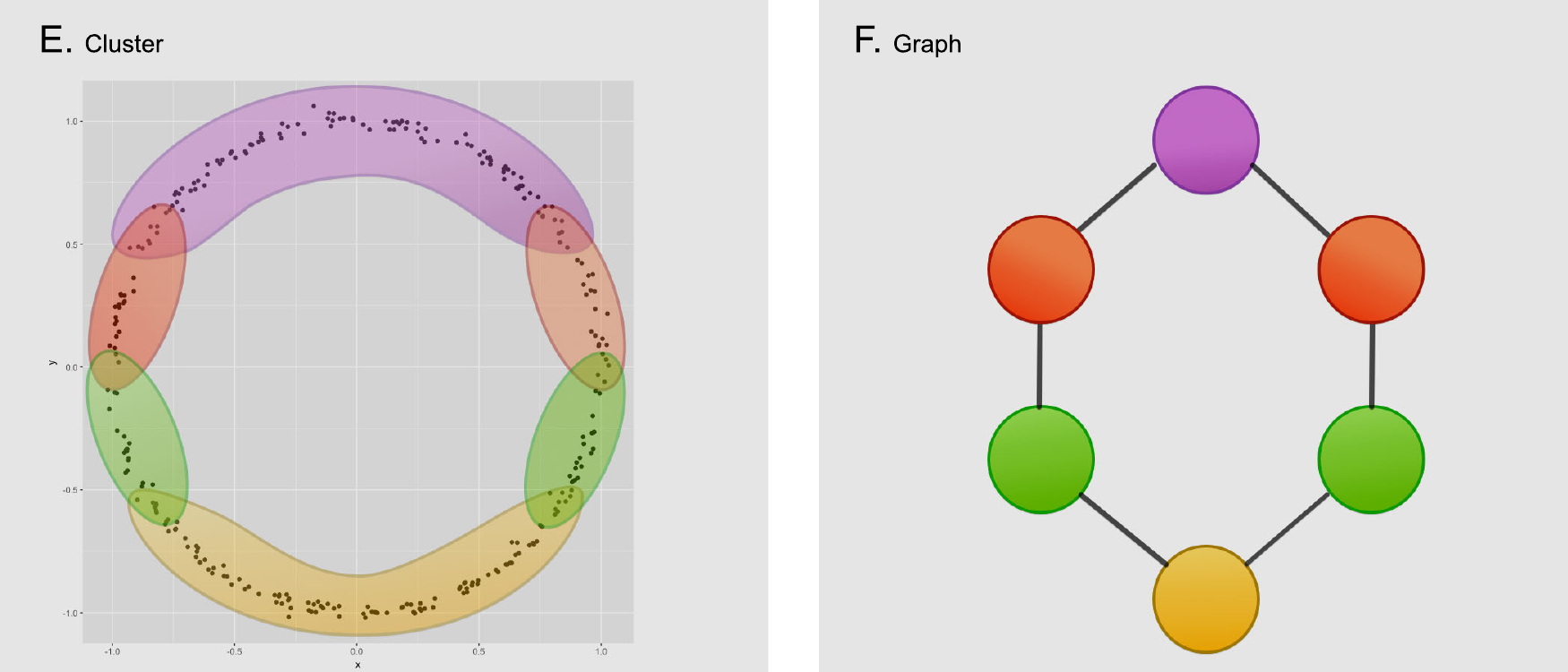}
    \caption{Inside every interval defined subset of data, we execute a clustering algorithm to detect isolated groups   of points (image E). Each of the resulting groups will correspond to a node on the output graph (image F). Notice how nodes are joined together by arches when their corresponding groups have points of the data set in common. }
  \label{fig:mapper3}
\end{figure}

The input of the method \textit{Mapper} is a collection of points with a notion of similarity and a filter, a function defined on the collection of points. The filter is used to define pieces that cover the collection of points. We apply a clustering algorithm to each piece to obtain a set of local clusters. These are the vertices of the graph. Edges are added to the graph in the following way: two local clusters are connected if they have points in common. Since each vertex and edge correspond to subcollection of points, we can consider the graph to have weights. Figs \ref{fig:mapper1} - \ref{fig:mapper3} give a visual road map through the \textit{Mapper} algorithm applied to a set sample from the unitary circle.

In this study, we implement TDA on disease data in Colombia to find topological characteristics that describe sociodemographic and spatiotemporal patterns.

For our each one of the 1,156 municipality we calculated an epidemic occurrence vector consisting of binary values for each week from 2007 to 2015. To construct each weekly entry, we executed a Kulldorf clustering procedure (as explained in the previous section) among all municipalities, but only with data from the given week. So a municipality \textit{k} will have 1 on a certain entry if in the corresponding week it suffered a malaria epidemic or 0 otherwise.\\

To this new sample of vectors we applied TDA, selecting the cosine distance as the similarity notion and the first and second principal components as the filter function, in order to search for significant clusters.

The output of the \textit{Mapper} algorithm is a graph where we select significant subgraphs by their size and other characteristics. To visualize the municipalities that appear in the subgraphs of the \textit{Mapper} output we make another graph were the nodes are the municipalities that appear in several nodes.

 \subsection{Ethnicity}

Case reports, collected by the national surveillance system and confirmed by laboratory, include components of age and ethnicity in the notification form. This filled form is required by law for every case reported in the country, and treatment to cure the disease is provided by the government for each case. Ethnicity and age are self-reported, and should be interpreted with caution (e.g. no genetic resistance can be inferred, for example). Three main ethnic groups were used in this paper: \textit{indigenous, Afrocolombian} and \textit{other} (with no ethnic denomination, ND).

 Data for age and ethnicity were displayed with descriptive statistics to develop patterns of malaria intensity by population for the clusters identified with TDA analysis, and to identify two patterns of intensity. First, the occupational hazard risk profile is characterized by a peak within a particular age-group and contains a pronounced sex difference \cite{da2012amazonian,alves2005asymptomatic}. Second, an endemic risk profile is characterized by intense malaria exposure at young ages and reduced malaria at later ages, due to overexposure \cite{da2012amazonian,alves2005asymptomatic}.

\section{Results}

\subsection{Clustering}

\begin{figure}
  \begin{center}
    \includegraphics{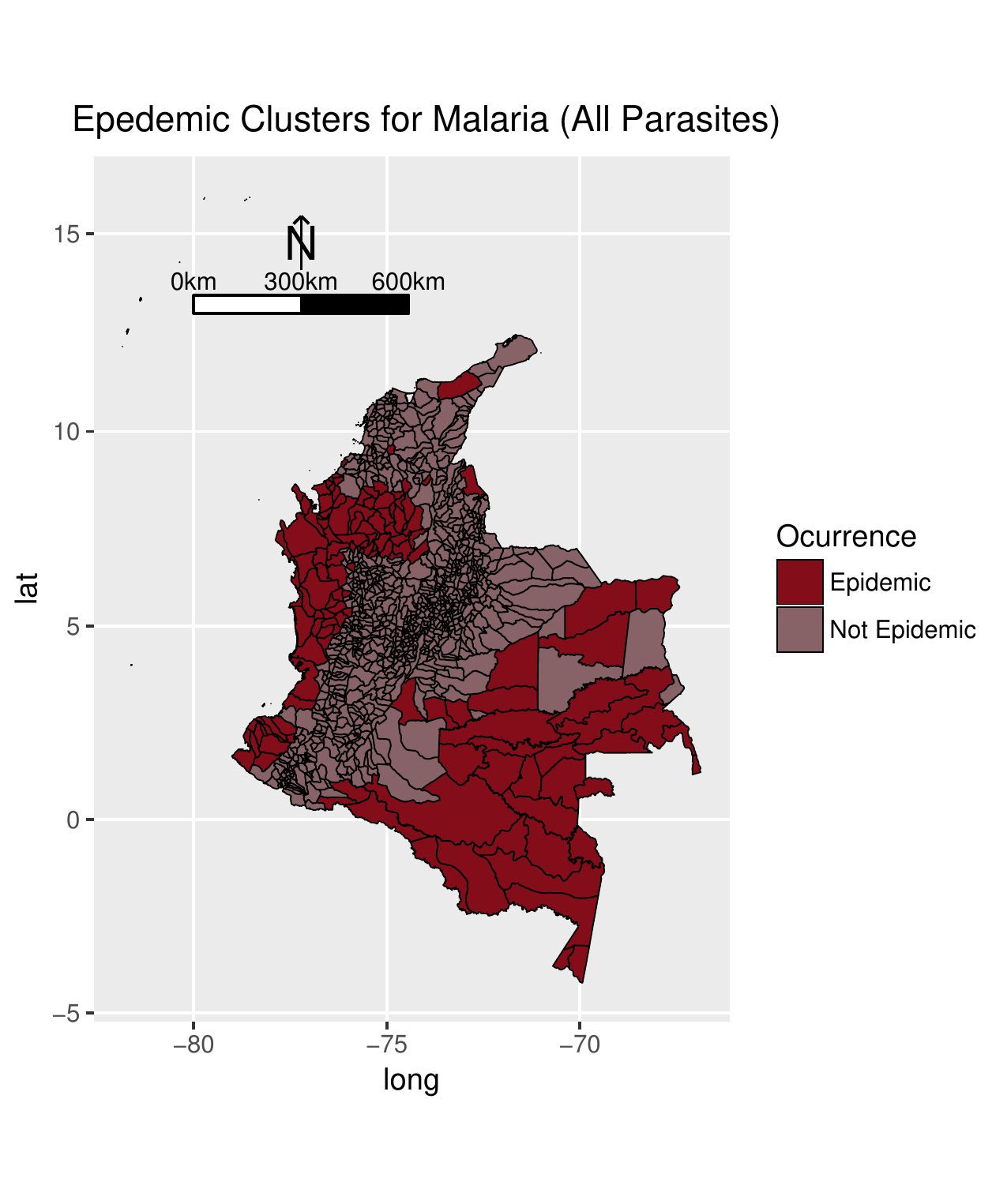}
  \end{center}  
  \caption[Significant outbreaks of malaria (all parasites) in Colombia from 2007-2015]{Significant outbreaks of malaria (all parasites) in Colombia from 2007-2015, calculated using the scan statistic developed by \cite{kulldorff1997} based on a likelihood ratio. The significance threshold parameter was calculated using a Bernoulli model where cases were simulated for each municipality, and taking the maximum value. The process was iterated many times and the distribution of the maximum values was calculated to determine the 95\% confidence interval. The method detects significant outbreaks of malarial infection along the Pacific Coast, the Magdalena river Basin, and the Amazon river Basin.}
  \label{clusters}
\end{figure}

\begin{figure}
  \begin{center}
    \includegraphics{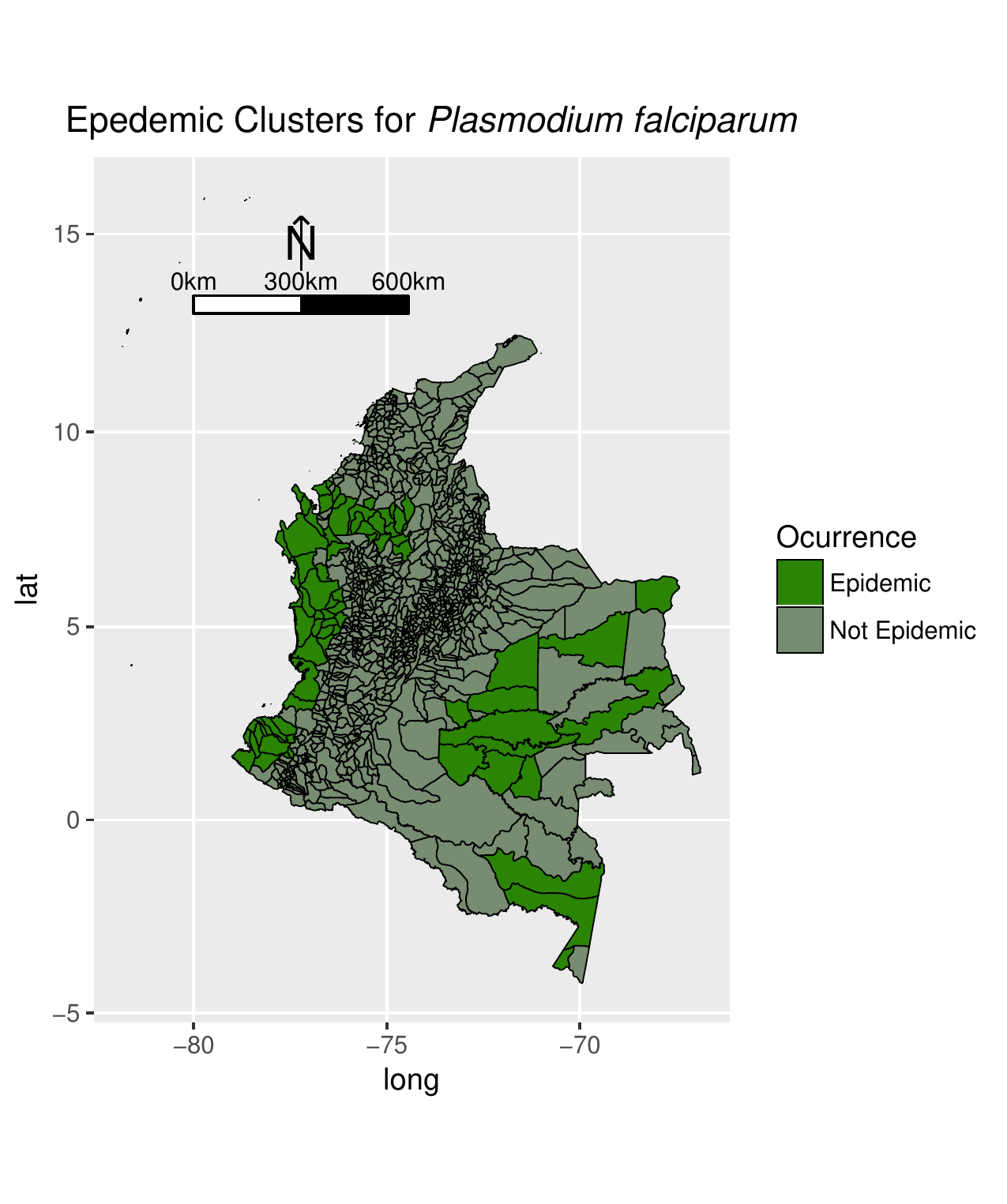}
  \end{center}  
  \caption[Significant outbreaks of \textit{Plasmodium falciparum} in Colombia from 2007-2015]{Significant outbreaks of \textit{Plasmodium falciparum} in Colombia from 2007-2015. The method used to find significant outbreaks is the same as described for Fig~\ref{clusters}. Significant clusters were observed in municipalities along the Pacific Coast, the border with Panama, and Northen Antioquia (the tertiary Cauca river Basin). The municipalities: Policarpa and Cumbitirá, in Nari\~{n}o appear to be a hidden cluster for this parasite, since they were not marked as epidemic when considering all malarial parasites. (Fig ~\ref{clusters})}
  
  \label{clustersfalc}
\end{figure}

\begin{figure}
  \begin{center}
    \includegraphics{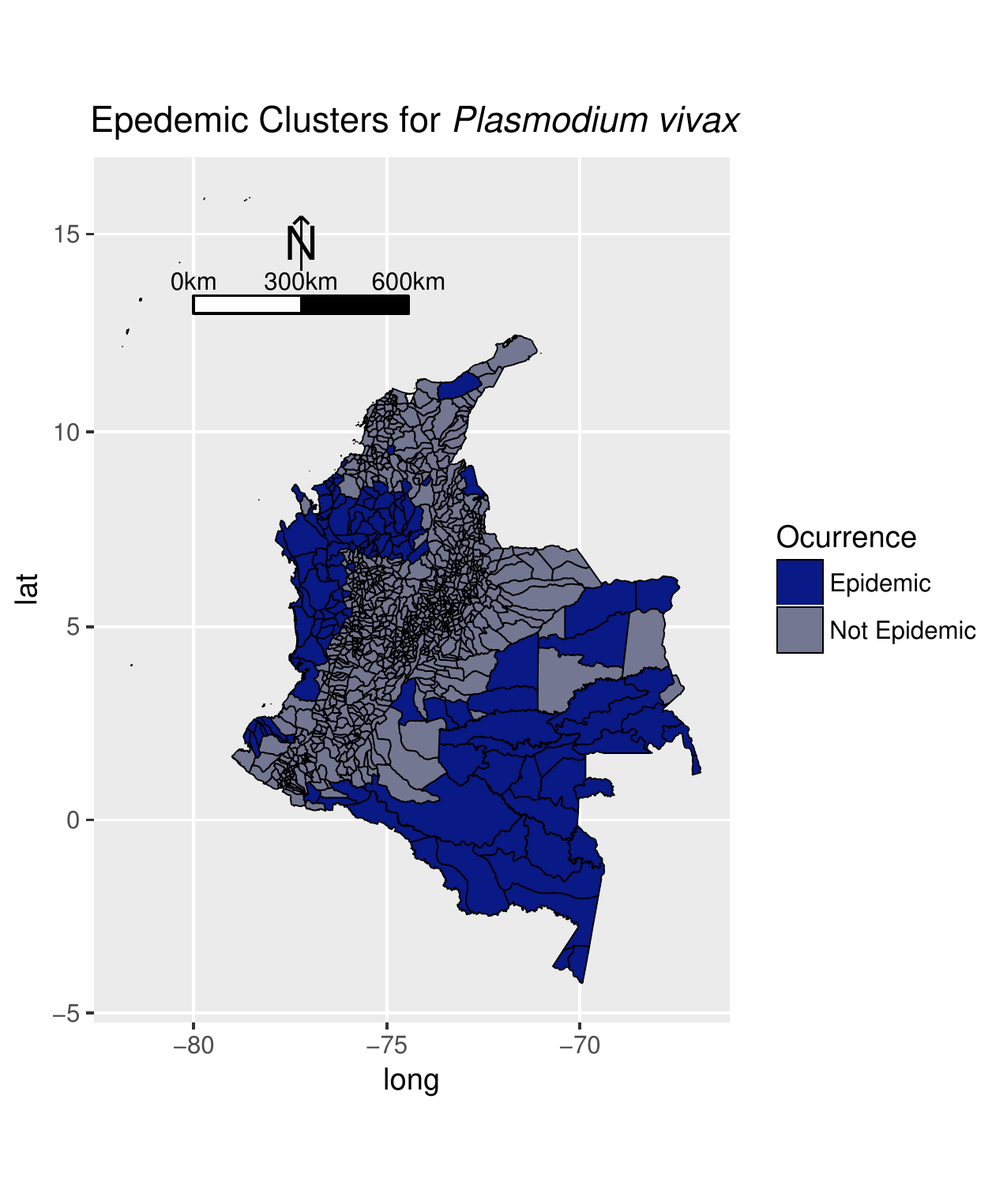}
  \end{center}
  \caption[Significant outbreaks of \textit{Plasmodium vivax} in Colombia from 2007-2015]{Significant outbreaks of \textit{Plasmodium vivax} in Colombia from 2007-2015. The method used to find significant outbreaks is the same as described for Fig~\ref{clusters}. Significant clusters were observed in municipalities of departments: Cordoba, Vichada and Antioquia. The municipality: Orito in Putumayo appears to be a hidden cluster for this parasite, since they weren't marked as epidemic when considering all malarian parasites. (Fig ~\ref{clusters})}
  
\label{clustersviv}
\end{figure}

We found an outbreak of \textit{Plasmodium vivax} malaria that comprised the Amazon Basin, including the departments of Amazonas, Caquetá, Meta, Guaviare, Putumayo and Nari\~no. Regions of Vichada, Chocó, Caldas and Antioquia also presented outbreaks, as did the region surrounding Barranquilla in the Caribbean. Singular clusters for this species where detected in parts of Putumayo. Significant outbreaks of \textit{Plasmodium falciparum} malaria in municipalities of Chocó, Risaralda, Antioquia, Nari\~no and Guaviare. Singular clusters for this species where detected in parts of Nari\~no. All significant outbreaks are highlighted in Fig~\ref{clusters}-\ref{clustersviv}.

\subsection{Synchronous epidemic visualization}

\begin{figure}
\hspace*{-2.25in}
\centering
  \includegraphics{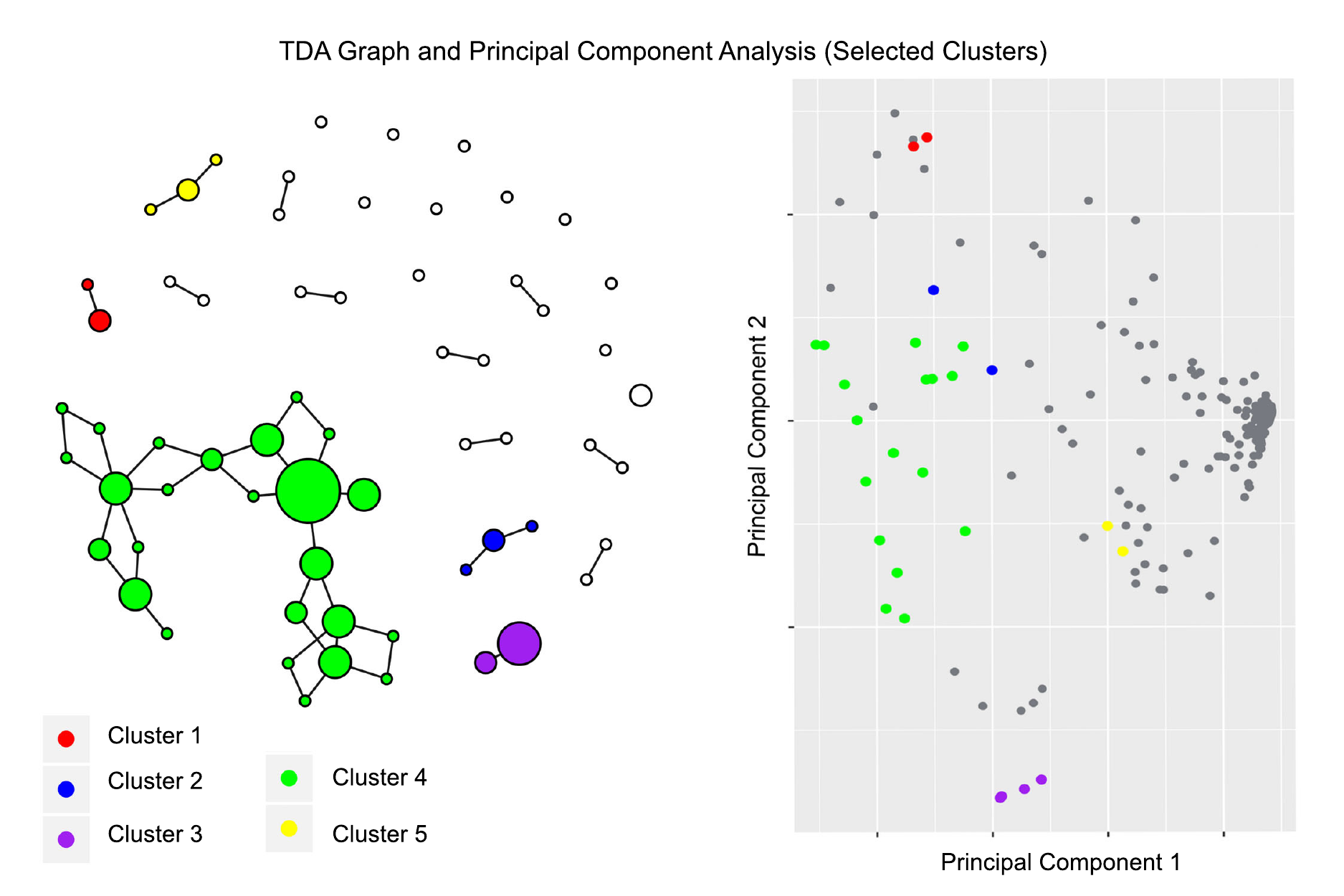}
    \caption{Graph constructed using TDA and principal component plot using the epidemic occurrence vectors, where selected groups have been highlighted. Highlighted groups where selected by size and high overall disease intensity (Clusters 1,2,5 have an average of 10\% disease intensity). Each cluster can be interpreted as a group municipalities with malaria incidence that have similar temporal behavior. }
  \label{res1_g}
\end{figure}

\begin{figure}
  \centering
  \includegraphics{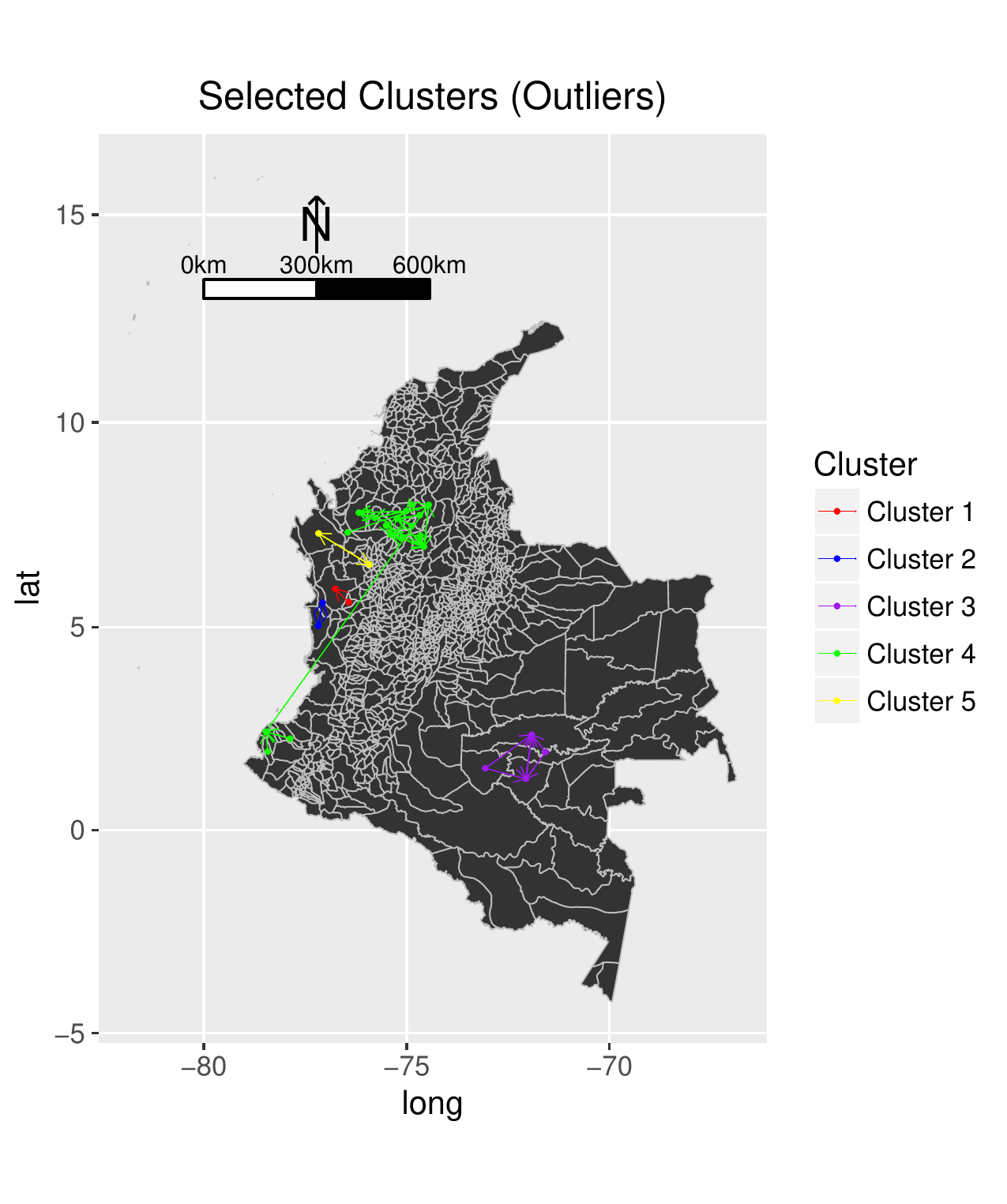}
    \caption{Selected municipalities by TDA over the Colombian territory. As expected, the clusters follow geographic pattern, since the time series where constructed using a Kulldorf procedure that detects clusters geographically. An unexpected result happens in cluster 4, where the cluster is divided by two major non adjacent regions: Antioquia and Nari\~{n}o. This means that the malaria epidemic follows the same time pattern among this two regions, even though they are geographically separate.}
  \label{tda_map}
\end{figure}

The TDA enabled us to find at least 5 groups of municipalities with similar behaviors (Fig \ref{res1_g} and Fig \ref{tda_map}), in particular, three of these group show  clusters with high overall disease intensity, defined as the number of cases with infants (age below 5 years) over the total number of cases. The geographic distribution of these clusters is as follows: two of this groups are  respectively concentrated on the Pacific and Caribbean coasts, showing a geographic relation among their municipalities, the remaining three groups have their members scattered around different parts of the country, including Chocó, Guania, Antioquia and Casanare.\\

In the TDA graph each node represents a group of municipalities. The size of each node will be proportionate to the number of municipalities in the group, and two nodes will have an arch between them if they have at least one municipality in common.

Now, it is possible to visualize the municipalities appearing in the TDA graph geographically, to get a notion of the part of Colombia this cluster represents.

Each of the colored small dots in the map corresponds to a municipality, contained in some node of the corresponding colored cluster. Notice there are arrows between some points in the map, they represent connections between the municipalities (these connections are not the same as  the arches between nodes in the graph). Before we mention how these connections are constructed, lets us recall what centrality means for a municipality:

Given a certain municipality, its centrality corresponds to the number of nodes in the graph it appears in. This means that if we remove a municipality with high centrality, it is very possible that the resulting graph will have less arches and in turn be disconnected.

Now, the connection scheme is as follows:
Only municipalities that appear in several nodes in the TDA graph can have a connection.
No municipality will be connected to itself.
Municipalities will be connected towards the municipalities in its node with the most centrality. Note that there could be municipalities with multiple outgoing connections.

We also identified five significant municipalities with high centrality in the TDA graph (Fig \ref{res1_g}) and are reported in table \ref{tda_table}. This municipalities are responsible for the connection among several nodes in their corresponding subgraphs appearing in overlapping zones of the selected TDA filter.

\begin{table}
\begin{adjustwidth}{-2.25in}{0in} 
\centering
\caption{
{ Selected central municipalities after executing TDA over the epidemic occurrence vectors. These are the municipalities responsible for the connectivity among their respective groups and subgraphs.}}
\begin{tabular}{rllrrr}
  \hline
Cluster & State Name & Locality & Rural Popu. & Urban Popu. & Total Popu. \\ 
  \hline
 1 & Chocó & Quibdó & 11752 & 101134 & 112886 \\ 
  2 & Chocó & Bajo Baudó & 13752 & 2623 & 16375 \\ 
   3 & Guaviare & San José del Guaviare & 19131 & 34863 & 53994 \\ 
  4 & Córdoba & Tierralta & 45895 & 32875 & 78770 \\ 
   5 & Antioquia & Santa Fe de Antioquia & 9267 & 13636 & 22903 \\ 
   \hline
\end{tabular}
\label{tda_table}
\end{adjustwidth}
\end{table}

\subsection{Ethnicity}
\begin{figure}
  \centering
  \includegraphics{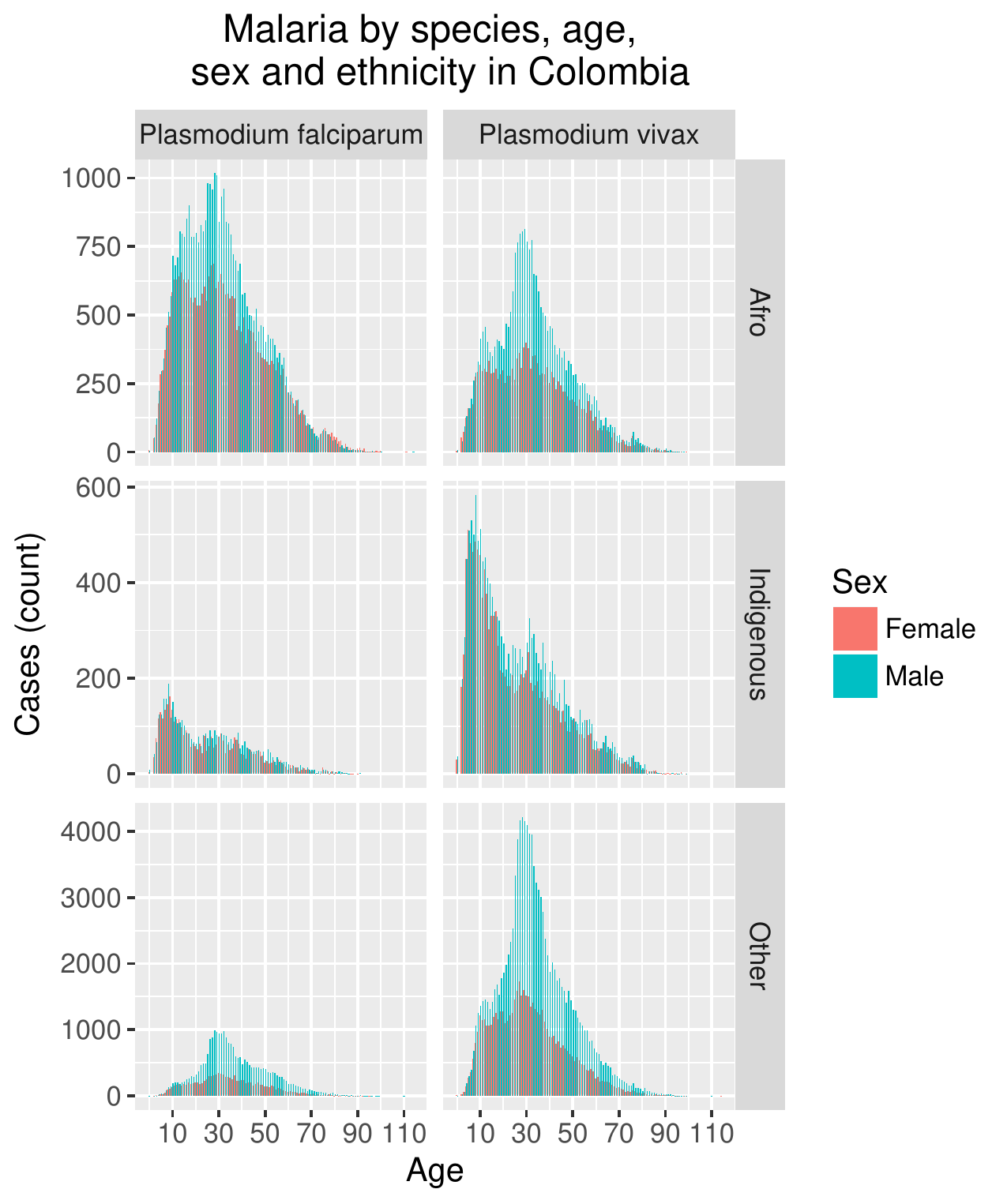}
    \caption{Malaria by parasite species, age, sex and ethnicity group of human cases in Colombia. The indigenous ethnic group shows a pattern of endemicity, with most cases being reported for the youngest ages, while people with no ethnic denomination and the Afrocolombian population show a pattern consistent with occupational hazard risk.}
  \label{bars_all}
\end{figure}

\begin{figure}
  \centering
  \includegraphics{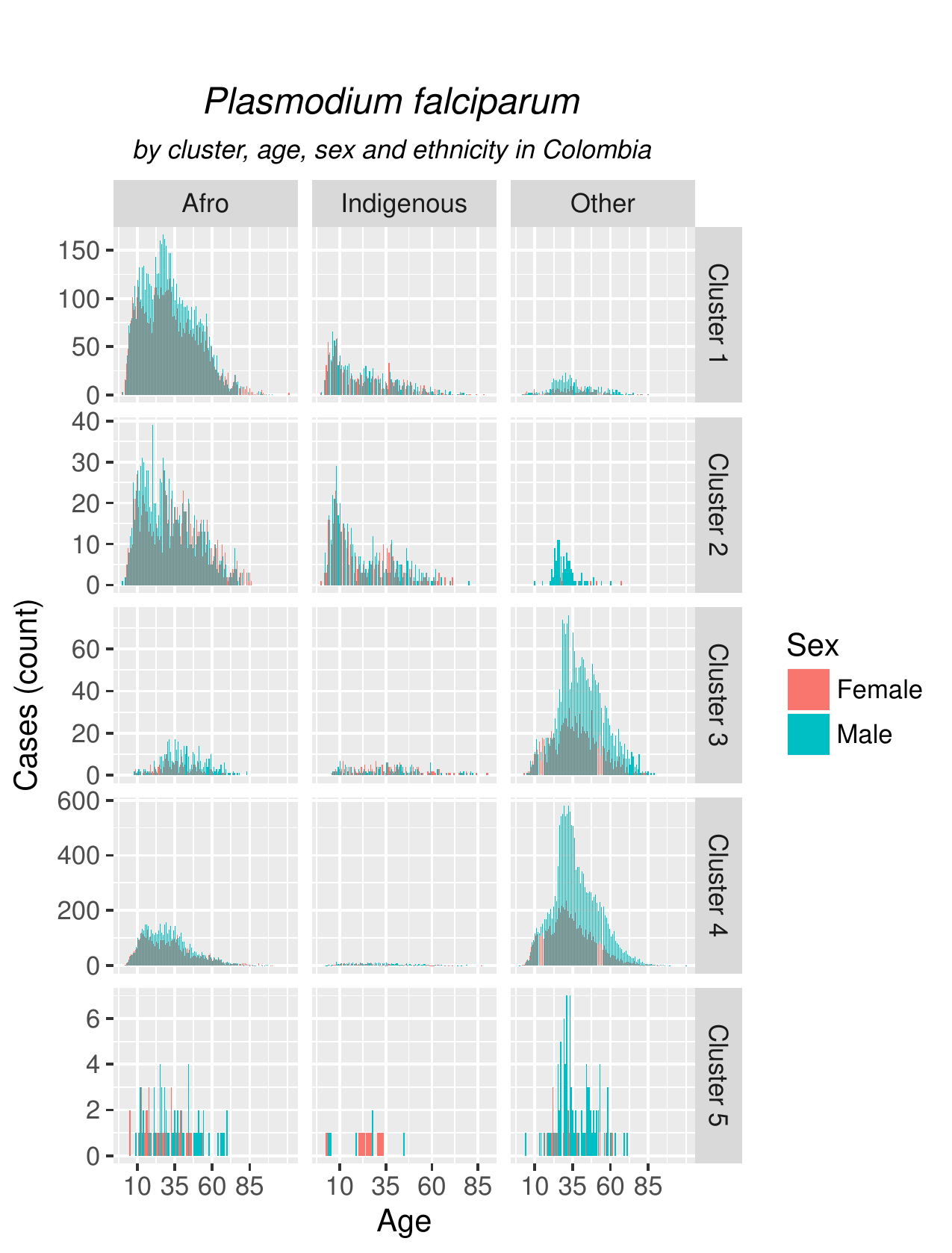}
    \caption{Malarial infection by \textit{Plasmodium falciparum} by cluster, age, sex and ethnicity in Colombia. Histograms for the indigenous population in clusters 1 and 2 suggest that these populations experience intense exposure to malarial infection. Similarly, histograms for the Afrocolombian population in clusters 1 and 2 suggest lower intensisty of malarial infection, cluster 1 experiencing more occupational hazard than cluster 2. Histograms for the population with no ethnic denomination in clusters 1, 2, and 4 suggest malarial infection is associated with occupational hazard. Clusters 3 and 5 have too few cases to infer either transmission intensity or occupational hazard.}
  \label{bars_falciparum_big}
\end{figure}

\begin{figure}
  \centering
  \includegraphics{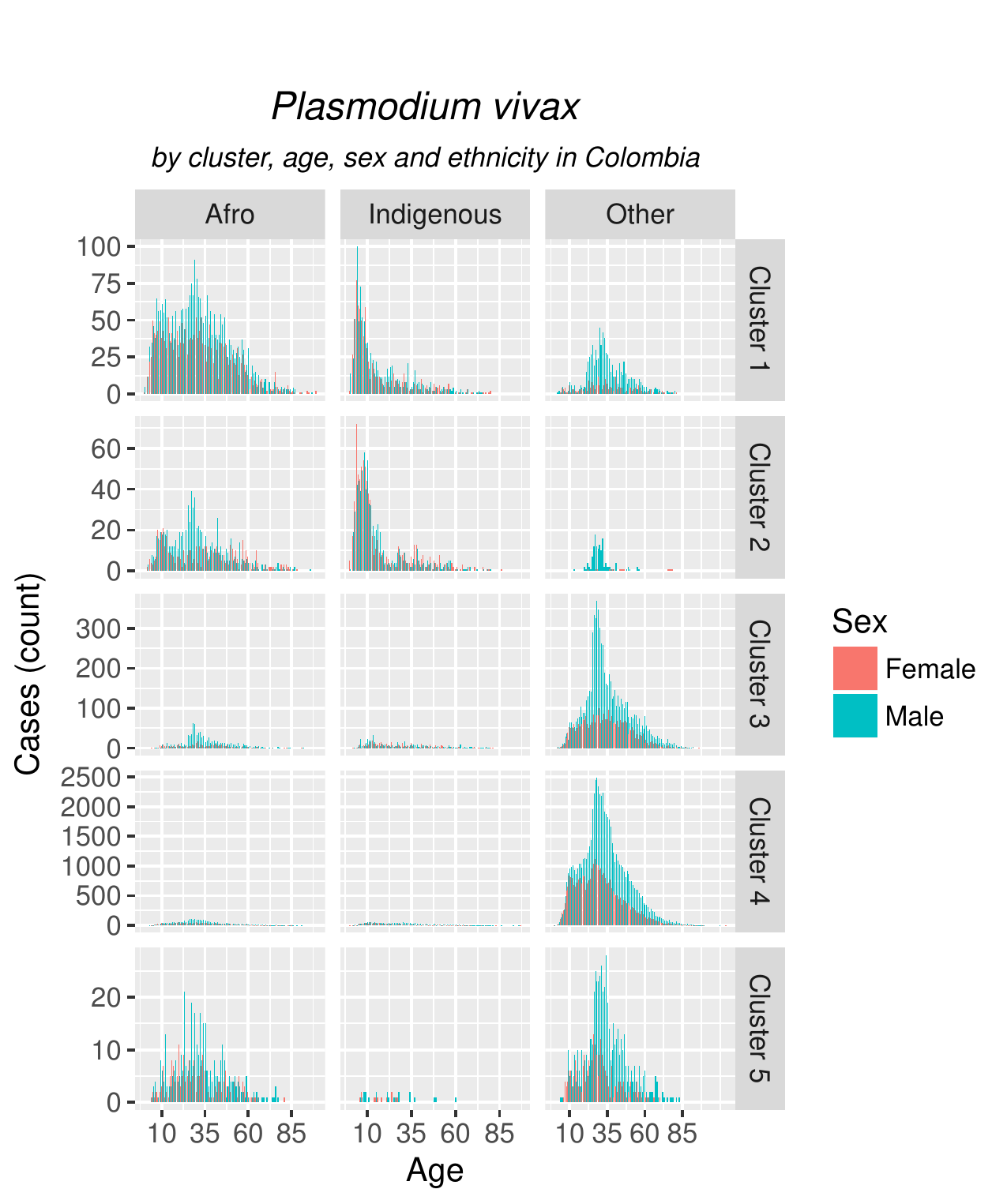}
    \caption{Malarial infection by \textit{Plasmodium vivax} by cluster, age, sex and ethnicity in Colombia. Histograms for the indigenous population in clusters 1, 2, and 3 suggest intense exposure to malarial infection among these populations, with cluster 2 experiencing the most intense exposure. Histograms for the Afrocolombian populations of clusters 1, 2, and 3 suggest some degree of occupational hazard transmission. The population with no ethnic denomination experiences malarial infection as an occupational hazard in all clusters except in 5. It is interesting that cluster 3 is the only cluster where an occupational hazard exists without another population experiencing endemic malaria in the same region.}
  \label{bars_vivax_big}
\end{figure}

Fig~\ref{bars_all}-\ref{bars_vivax_big} show the histograms of age reports of malaria by ethnicity and cluster. Two distinctive patterns consistently appearing throughout different regions of Colombia. First, an endemic profile risk (higher density of cases at young ages) was observed for the indigenous populations of clusters 1 and 2. Second, malarial infection suggesting occupational hazard and intensity of infection among the Afrocolombian populations of clusters 1 and 2. Occupational hazard is consistently described for the population with no ethnic denomination across clusters 1-4. In all cases, within the same cluster, we find both endemic and occupational hazard infection patterns across populations segregated by ethnicity. In a histogram of case reports by age and sex, an occupational risk hazard has a unique and characteristic signature: one age class, typically for only one sex, presents an outstanding number of cases compared to any other age class. In the case of malaria in Colombia, we observed that men with no ethnic denomination of ages 20-25 were contracting malaria far more often than any other class. From this simple observation, we inferred the following: first, men of this age class were engaging in activities that posed a risk of contracting the disease. Second, women were not engaging in this activity, nor were men in other age classes. Third, there was no household transmission, since infected men were not infecting other members of their family once they ceased to engage in the risky activity.

\section{Conclusions}

Malaria in Colombia was characterized by a different intensity, connectivity and segregation in each region. While there was a general pattern of risk throughout the country associated with occupational hazard, some populations experienced intense malaria exposure in endemic pockets. Understanding the interaction of such pockets is fundamental for designing appropriate malarial control strategies. Here we have produced a systematic approach that analysis of malaria under three dimensions.

Colombia experienced a generalized malaria outbreak in the Amazon region for the period studied. We found that there was little connectivity among the municipalities that composed the Amazon region, and that this outbreak was spatially connected to the Cauca Basin in Northen Antioquia. In the clusters connected to the Amazon region, where there was relatively high degree of cultural diversity, indigenous populations experienced malaria in endemic patterns, contrary to the risk of the ND population for both the region and for the country.

The Cauca basin was characterized by different pattern: the Afrocolombian population experienced a segregated exposure to \textit{Plasmodium falciparum} in a way that no other ethnic group did. The pattern observed for most of the country is not consistent across the Pacific region, where \textit{Plasmodium falciparum} also persisted at a relatively higher prevalence than in the rest of the country in comparison to \textit{Plasmodium vivax}. Most interestingly, the Cauca Basin region contained two different populations that lived in pockets of endemicity, while it is also synchronous with other regions, and furthermore it is part of a region where the scan statistic algorithm detected an outbreak.

Our findings have potential implications for malarial infection control. First, we found that malaria in Colombia did present different, isolated pockets with distinctive epidemic characteristics. The magnitude of such differences in epidemic characteristics is relevant in studying the pressure of anti-malarials upon the parasite, since the emergence of resistance has been reported in the country. We found that \textit{Plasmodium falciparum} was particularly acute among the Afrocolombian population of the Cauca Basin and the Pacific region. However, in the Cauca Basin, it constituted an isolated outbreak, while in the Pacific, the outbreak was dispersed among both the Afrocolombian and the indigenous populations. Different parasite loads among ethnically and culturally distinct populations constitute the quintessential mechanism of selective pressures that are ideal for the evolution of parasites. The diversity of epidemic characteristics of malarial infection among the subpopulations of Colombia account for an ideal environment for parasite evolution, where plasmodia persist under different pressures of asymptomatic individuals, susceptible classes of ethnically distinct populations, and public health interventions using different anti-malarial strategies. Such diversity provides the necessary conditions, acting as isolated experiments, and then sharing ``successful" results, for the emergence of resistant parasites.  

Second, the patterns of endemicity observed in these populations suggested that prevention efforts should be population specific, and vary according to the epidemic characteristics exhibited by the parasite in the targeted population. Therapeutic failures have been suggested to be correlated with high intestinal parasite loads \cite{blair2002resistance}. The effectiveness of bed nets has been reported to be low among populations that experience intense malaria exposure \cite{snow1996need}. We have identified populations that experienced malaria endemicity, where prevention efforts focused on the distribution of bed nets. Our findings, combined with previous knowledge suggest that public health interventions should integrate two aspects: 1)  Diagnostic and treatment of asymptomatic malaria; and 2)  Diagnostic and treatment of intestinal parasites (to reduce therapeutic failure).

Third, prevention strategies focusing on populations with endemic malaria would yield a reduction of occupational hazard malaria, since the occupational hazard is associated to visiting locations where malaria persists endemically. 

\clearpage

\section{Additional Figures}

\begin{figure}
  \begin{center}
    \includegraphics{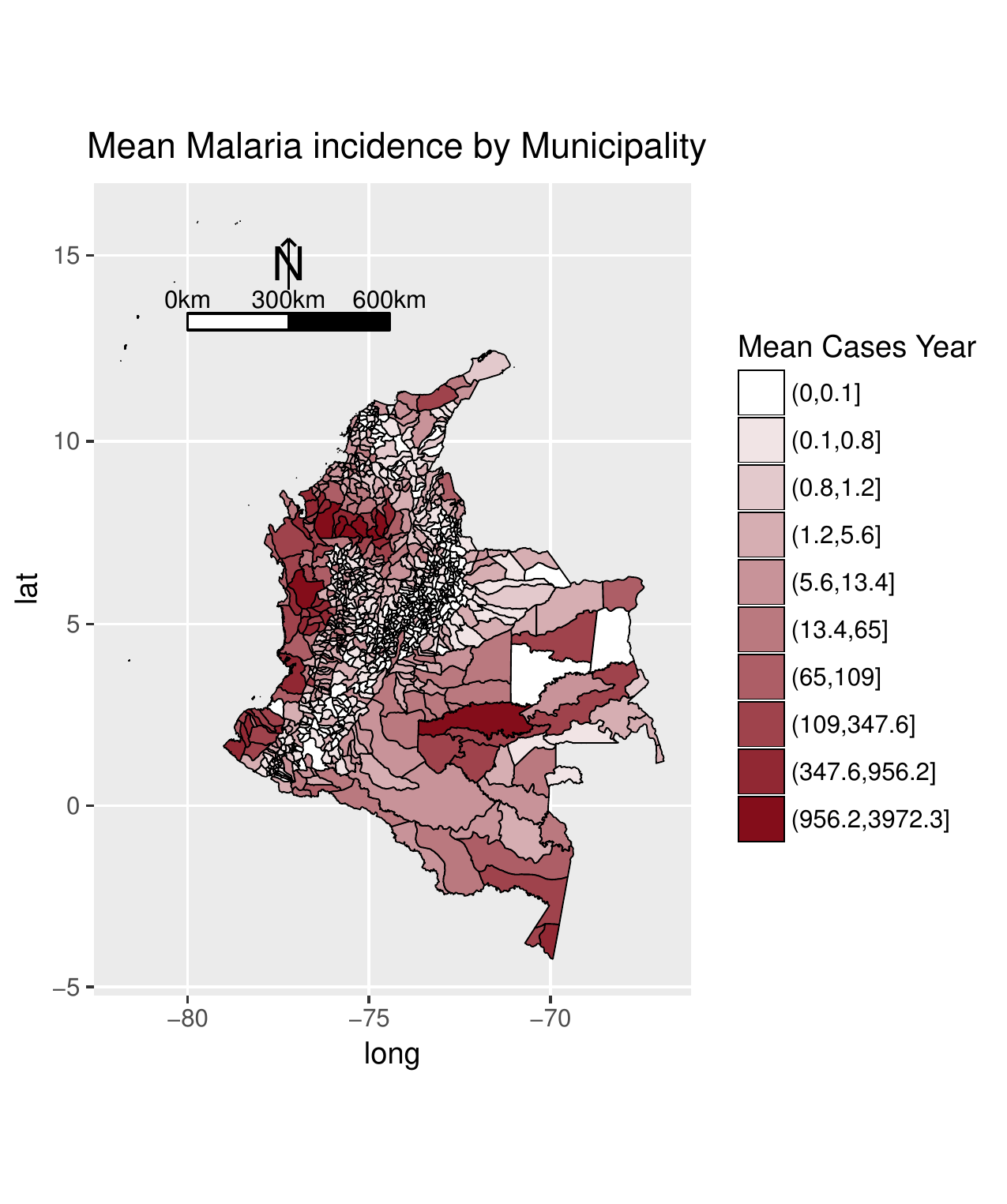}
  \end{center}
  \caption{Malarial incidence for both species in Colombia from 2007-2015. Intervals where constructed using the Jenks procedure \cite{jenks1971error}}
\end{figure}

\begin{figure}
  \begin{center}
    \includegraphics {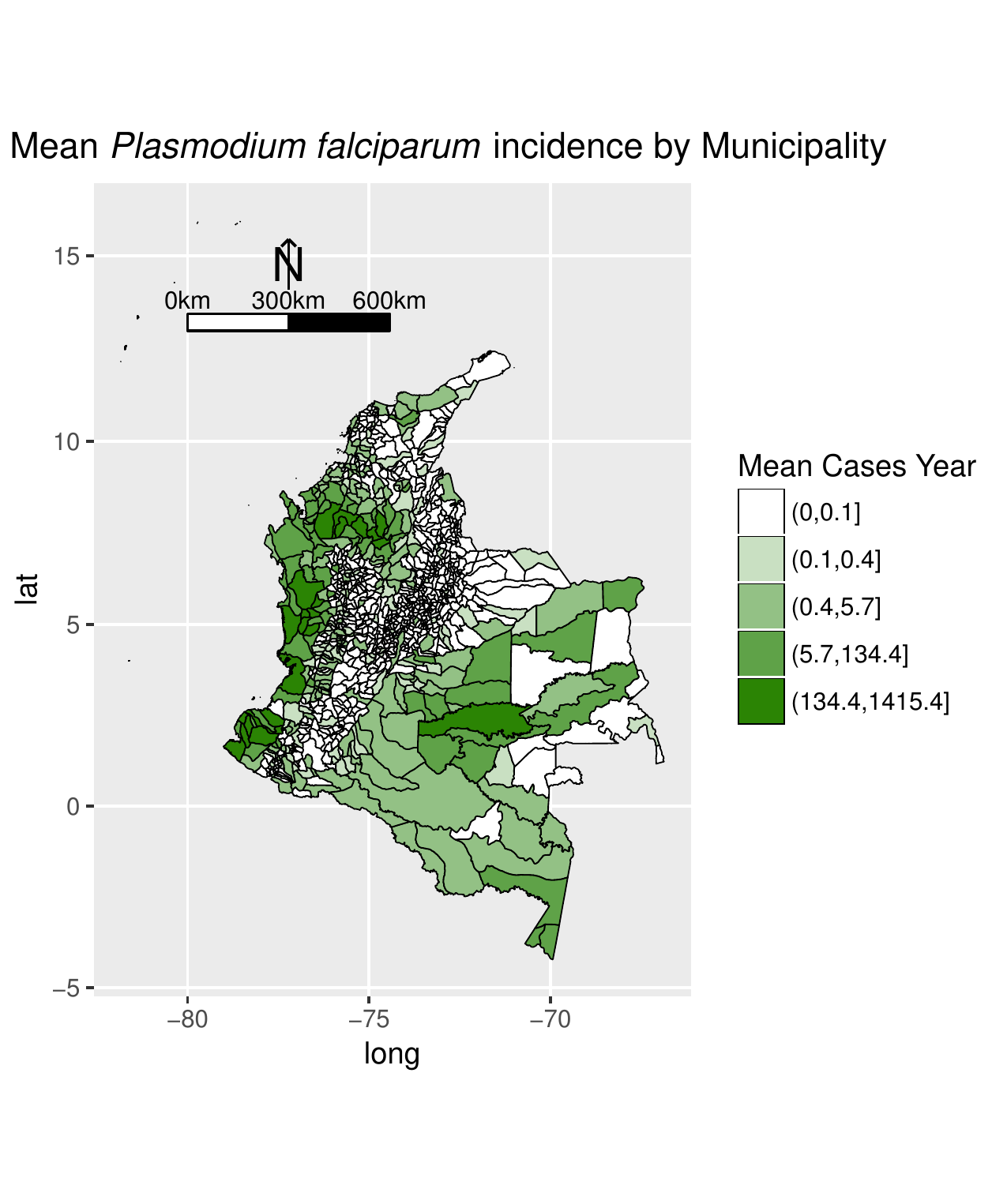}
  \end{center}  
  \caption{\textit{Plasmodium falciparum} incidence in Colombia from 2007-2015. Intervals where constructed using the Jenks procedure \cite{jenks1971error}}
\end{figure}

\begin{figure}
  \begin{center}
    \includegraphics{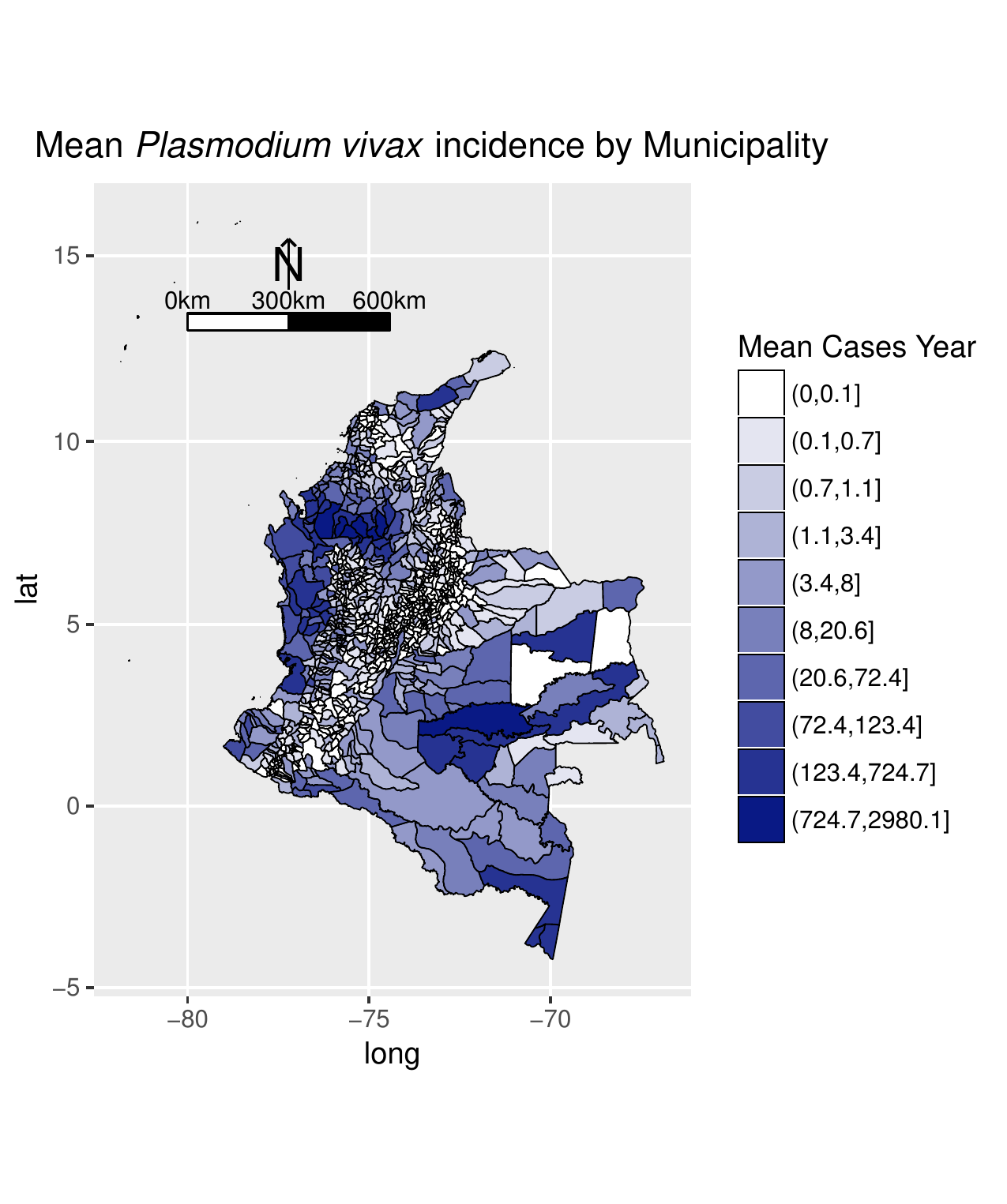}
  \end{center}
  \caption{\textit{Plasmodium vivax} incidence in Colombia from 2007-2015. Intervals where constructed using the Jenks procedure \cite{jenks1971error}}
\end{figure}

\section*{Acknowledgments}

The first author would like to thank Stanford University, the Zaffaroni family, and the Morrison Institute for their financial support. 
The second author acknowledges and thanks the financial support of the  grant \textit{ P12.160422.004/01- FAPA ANDRES ANGEL} from Vicedecanatura de Investigaciones de la Facultad de Ciencias de la Universidad de
los Andes, Colombia. The third author acknowledges and thanks the financial support of the grant \textit{Proyecto Semilla 2017-1} from Vicedecanatura de Investigaciones de la Facultad de Ciencias de la Universidad de
los Andes, Colombia

%
%
%


\end{document}